\documentclass[aps,prl,twocolumn,groupedaddress,amsmath,amssymb,preprintnumbers]{revtex4}
\usepackage{color,slashed,graphicx,changes}
\usepackage[utf8]{inputenc}

\usepackage{hyperref}

\usepackage{soul,xcolor}
\usepackage{graphicx}
\usepackage{amsmath}
\usepackage{amssymb}
\usepackage{verbatim}
\usepackage{mathrsfs}
\usepackage{array}
\usepackage{layout}
\usepackage{textcomp}
\usepackage{latexsym}
\usepackage{slashed}
\usepackage{rotating}
\usepackage{hyperref}
\usepackage{booktabs}
\usepackage{relsize}
\usepackage{multirow}

\newcommand\cyansout{\bgroup\markoverwith{\textcolor{cyan}{\rule[0.5ex]{2pt}{0.4pt}}}\ULon}
\newcommand\blueout{\bgroup\markoverwith{\textcolor{blue}{\rule[0.5ex]{2pt}{0.4pt}}}\ULon}

\hypersetup{
	colorlinks = true,
	linkcolor = blue,
	citecolor = magenta
} 
\newcommand{\Eq}{&=&}

\newcommand{\bs}[1]{{\boldsymbol{#1}}}
\newcommand{\hs}[1]{{\hspace{#1}}}
\newcommand{\vs}[1]{{\vspace{#1}}}
\newcommand{\tf}[1]{{\textsf{#1}^{}}}
\newcommand{\tx}[1]{{\text{#1}^{}}}
\newcommand{\nn}{\nonumber\\}

\newcommand{\dd}{{\mathrm{d}}}

\newcommand{\UoneX}{\tx{U}(1)_{\tf{L}_\mu - \tf{L}_\tau}}
\newcommand{\BKnunu}{B^+ \!\to K^+\nu\bar{\nu}}

\begin{document}


\title{Recent $\bs{\BKnunu}$ excess and muon $\bs{g - 2}$ illuminating 
\\ light dark sector with Higgs portal}


\author{Shu-Yu\,\,Ho$^{a}$} \thanks{phyhunter@kias.re.kr}

\author{Jongkuk\,\,Kim$^{a,b}$}
\thanks{jkkim@kias.re.kr, {\it corresponding author}}

\author{Pyungwon\,\,Ko$^{a}$}
\thanks{pko@kias.re.kr}

\affiliation{$^a$ Korea Institute for Advanced Study, Seoul 02455, Republic of Korea}
\affiliation{$^b$ Department of Physics, Chung-Ang University, Seoul 06974, Korea}

\preprint{KIAS-P24003}

\begin{abstract}
The Belle II collaboration recently announced that they observed the $\BKnunu$ decay process for the first time. However, their result encounters a $2.7{}^{}\sigma$ deviation from the Standard Model (SM) calculation.\,\,Additionally, Fermilab released new data on muon $g-2$ away from the SM expectation with $5.1{}^{}\sigma$.\,\,In this letter, we study the simplest UV-complete $\UoneX$-charged complex scalar Dark Matter (DM) model.\,\,Thanks to the existence of light dark Higgs boson and light dark photon, we can explain the observed relic density of DM and resolve the results reported by both Belle II and Fermilab experiments simultaneously.\,\,As a byproduct, the Hubble tension can be alleviated by taking $\Delta N_\tf{eff} \simeq 0.3$ induced by the light dark photon, which could be tested by CMB stage-4 and new NA64 experimental data in the near future.\,\,In addition, our light DM mass is highly testified by future data released by Belle II and CMB stage-4. 
\end{abstract}


\maketitle

\section{Introduction}
Recently, the Belle II collaboration has found the first evidence of the $\BKnunu$ rare decay predicted by the SM~\cite{Belle-II:2023esi}.\,\,The decay branching fraction was measured with two different techniques\,:\,Hadronic-tagged and inclusive-tagged analyses.\,\,The combined result is given by
\begin{eqnarray}
{\cal B}\!\left( \BKnunu \right)_\tf{exp} 
=\,  
\left(2.3 \pm 0.7 \right) \times 10^{-5}
~.
\end{eqnarray}
On the other hand, it is known that the decay branching fraction of $\BKnunu$ is theoretically clean, 
and calculated with high accuracy in the SM~\cite{Parrott:2022zte}:
\begin{eqnarray}
{\cal B}\!\left( \BKnunu \right)_\tf{SM} 
=\, \left( 4.97 \pm 0.37 \right) \times 10^{-6}
~.
\end{eqnarray}
Comparing the difference between these two results, the Belle II measurement has a $2.7{}^{}\sigma$ deviation from the SM prediction.\,\,This deviation is not yet statistically significant enough to be interpreted as Beyond Standard Model (BSM) signals.\,\,However, it would be worthwhile to entertain the possibility of accommodating it in some well-motivated BSM models with light dark sectors.

Independent of $\BKnunu$,  the Fermilab Muon $g-2$ Collaboration released a new measurement of the muon magnetic moment in 2023~\cite{Muong-2:2023cdq}:
\begin{eqnarray}
\Delta a_\mu 
\,=\, 
a^\tf{exp}_\mu - a^\tf{SM}_\mu 
\,=\, 
\left(249 \pm 48 \right) \times 10^{-11}
~. 
\end{eqnarray}
This tension has increased from $4.2{}^{}\sigma$ released in 2021 to $5.1{}^{}\sigma$.\,\,However, it seems to reduce to $1.6{}^{}\sigma$ by taking into account the new lattice QCD result from BMW Collaboration~\cite{Borsanyi:2020mff}.\,\,Ongoing efforts are made to understand the discrepancy in the results obtained by two different theoretical approaches.
In this work, we will consider $5.1 \sigma$ is real to be explained by new physics BSM.
In the appendix, we consider another benchmark point $(m_{Z'}=10{\rm MeV}, g_X=10^{-4})$  where no discrepancy exists between experimental results and SM predictions for the muon $(g-2)$ including the error bars from prospective experiments.
This new BP can substantially relax the Hubble tension and be consistent  with current $\Delta a_\mu$ data within $1\sigma$ C. L.
For more details, please see the appendix.

If the mild excess in $B^+ \rightarrow K^+ \nu \bar{\nu}$ is real and to be interpreted in some new physics framework, a light dark sector would be most 
compelling scenario. However, light thermal WIMP below a few GeV is subject to stringent constraints from CMB and BBN. This constraint is usually evaded 
employing the forbidden dark matter scenario or the DM $p$-wave annihilation. 
Another new possibility which we would take in this work is to assume that 
the light mediators mostly decay into a pair of dark matter and dark photon, 
the latter of which decays to a pair of  neutrinos. 
Assuming this $Z'$ couple to both dark matter and neutrinos, we may be able to realize light WIMP for the Belle II excess without conflict with the CMB and BBN constraints.

We shall choose $\UoneX$ gauge extensions of the SM with complex scalar DM model, and demonstrate that it is indeed possible to interpret the Belle II excess in terms of light dark sector.
This Model can explain both $\Delta a_\mu$ and thermal DM if the vector boson $(V)$
mass in the Proca Lagrangian is near the resonance region, where $m^{}_V \simeq 2{}^{}m^{}_\tf{DM}$\,\cite{Foldenauer:2018zrz, Holst:2021lzm,Drees:2021rsg}.\,\,
However, it was emphasized in Ref.\,\cite{Baek:2022ozm} that this fine-tuned tight 
mass condition can be relaxed if we promote these models with the Abelian Higgs 
mechanism for the vector boson mass.  

In this letter, we fully take into account of dark 
Higgs boson in the whole analyses (See Ref.\,\cite{Ko:2018qxz} for reviews on 
the roles of dark Higgs boson in (astro) particle physics and cosmology).\,\,
Remarkably, we can explain the $\BKnunu$ excess through either two- or three-body decay accompanied by the dark Higgs boson with a nonzero vacuum expectation value (VEV) or DM particle, respectively.


\section{Model Overview}

A simple extension of the SM to account for the $\Delta a_\mu\!$ and $\BKnunu$ excess is the gauged $\UoneX\!$ model (known to be anomaly-free without adding extra chiral fermions~\cite{He:1990pn,He:1991qd}) including a dark Higgs boson.\,\,In this scenario, the 
$B^+ \!\!\to \! K^+ \!\nu\bar{\nu}$ excess can be addressed by the dark Higgs boson (or its dark decay product) that is interpreted as the missing energy akin to neutrinos.\,\,On the other 
hand, the $\Delta a_\mu\!$ can be explained by the dark $\UoneX\!$ gauge boson with a mass generated by the Higgs mechanism.\,\,To account for DM, we further introduce a complex scalar  charged under the $\UoneX\!$ symmetry which, as we will discuss soon, would mainly annihilate into the dark gauge bosons via the dark Higgs portal.\,\,Then these dark gauge bosons will eventually decay into neutrinos if they are lighter than the muon, which is the case we shall consider here.\,\,This is in contrast to the real singlet scalar DM via the Higgs portal where the final states of the DM annihilation are electrically charged SM particles~\cite{Silveira:1985rk,He:2009yd,He:2010nt}.

With the above model setup, the $\UoneX\!$ charge assignment for the relevant SM particles and new particles are given as follows
\begin{eqnarray}
\hs{-0.4cm}
\widehat{{\cal Q}}^{}_{\tf{L}_\mu - \tf{L}_\tau}
\big({}^{}\nu^{}_\mu, \nu^{}_\tau, \mu, \tau, X, \Phi \big)
=
\big({}^{}1, -1, 1, -1, {\cal Q}^{}_X, {\cal Q}^{}_\Phi \big)
\,,
\end{eqnarray}
where $X$ is the singlet scalar DM, and $\Phi = \frac{1}{\sqrt{2}} \left(\upsilon^{}_\Phi + \phi  \right)$ is the dark Higgs with a nonvanishing VEV $\upsilon^{}_\Phi$ that breaks the $\UoneX\!$ gauge symmetry spontaneously.\,\,To ensure the DM to be stable or very long-lived $(\tau^{}_X \gtrsim 10^{26}\,\tx{sec})$, here we assume that ${\cal Q}^{}_X \!= 1$ and ${\cal Q}^{}_\Phi$ in such a way that there are no gauge invariant operators up to dim-5 that would make the DM decay into the SM particles~\cite{Baek:2013qwa,Ko:2022kvl}.

Given this particle charge assignment, the renormalizable and gauge invariant Lagrangian is given by
\begin{eqnarray}
	{\cal L} 
	\,\Eq\,
	|{\cal D}_{\rho} \Phi|^2 + |{\cal D}_{\rho} X|^2
	- \tfrac{1}{4} 
	\big( \partial^{}_\rho Z'_\omega - \partial^{}_\omega Z'_\rho \big)^{\! 2}
	- m_X^2 |X|^2
	\nn
	&&\,
	- \,g^{}_X 
	\Big({}^{}
	\bar{\mu} \gamma^\rho\mu - \bar{\tau} \gamma^\rho\tau 
	+ \bar{\nu}^{}_{L\mu} \gamma^\rho \nu^{}_{L\mu} \!
	- \bar{\nu}^{}_{L\tau} \gamma^\rho \nu^{}_{L\tau}
	\Big)
	Z_\rho'
	\nn 
	&&\,
	- \,\lambda_{\Phi\!X} |X|^2 \Big( |\Phi |^2 - \tfrac{1}{2} \upsilon_\Phi^2 \Big) 
	- \lambda_{H\!X} |X|^2 \Big( |{\cal H}|^2 - \tfrac{1}{2} \upsilon_H^2 \Big) 
	\nn
	&&\,
	- \,\lambda_{\Phi\!H} 
	\Big( |\Phi |^2 - \tfrac{1}{2} \upsilon_\Phi^2 \Big)
	\Big( |{\cal H}|^2  - \tfrac{1}{2} \upsilon_H^2 \Big)  
	+\cdots
	~,
\end{eqnarray}
where ${\cal D}^{}_\rho \! = \partial^{}_\rho + i g^{}_X \widehat{{\cal Q}}^{}_{\tf{L}_\mu - \tf{L}_\tau} \! Z'_\rho$ is the covariant derivative with $g^{}_X\!$ denoting the dark gauge coupling and $\!Z'\!$ being the dark gauge boson with mass $m^{}_{Z'} \! = g^{}_X |{\cal Q}^{}_\Phi| \upsilon^{}_\Phi$, $m^{}_X\!$ is the mass of DM, and ${\cal H} = \frac{1}{\sqrt{2}} \left(0,\,\upsilon^{}_H + h \right)^{\hs{-0.05cm}\tf{T}}$ is the SM Higgs doublet (in the unitary gauge) with the VEV $\upsilon^{}_H \simeq 246.22\,\tx{GeV}$.\,\,From now on, we call $Z'$ as dark photon because it couples to the DM.\,\,Note that the kinetic mixing between $Z'$ and the usual photon arises through $\ell = \mu,\tau$ loops\,;\,$-\,\epsilon{}^{}{}^{}e{}^{}{}^{}\bar{\ell}\,\gamma^\rho \ell{}^{}Z'_\rho$ with $\epsilon \simeq -\,g^{}_X/ 70$~\cite{Escudero:2019gzq}.\,\,In this Lagrangian density, the DM relic abundance is determined by the $\lambda_{\Phi\!X}, \lambda_{H\!X}$, and $g^{}_X\!$ couplings.\,\,However, we will take $\lambda_{H\!X} \!= 0$ for simplicity.\,\,The $\lambda_{\Phi\!H}\!$ coupling allows the CP-even neutral components of ${\Phi}$ and ${\cal H}$, $\phi$ and $h$, respectively, to mix after electroweak and $\UoneX\!$ symmetry breakings.\,\,The dark Higgs (SM-like Higgs) boson in the mass eigenstate is denoted as $H_1(H_2)$, where $H_1 = \phi \cos\theta - h\sin \theta$ and $H_2 = \phi \sin\theta + h \cos\theta$ with $\theta$ being the mixing angle.\,\,In this work, we will assume that the mass of $H_1\!$ is smaller than that of $H_2$, $m^{}_{H_1} \!< m^{}_{H_2} \simeq 125\,\tx{GeV}$.


\section{Muon $\bs{g-2}$ \& Hubble tension}

In the $\UoneX\!$ model, the massive dark photon $Z'$ provides an additional contribution to the muon magnetic moment via the vertex correction~\cite{Baek:2001kca,Baek:2008nz}.\,\,Taking $g^{}_X\! \sim 10^{-4}$ and $m^{}_{Z'} \!< m^{}_\mu$, one can alleviate the discrepancy in $\Delta a_\mu$.\,\,The MeV-scale $Z'$ can produce substantial entropy and energy via the $Z'\to \nu\bar{\nu}$ decay process, which would spoil the successful predictions of the Big Bang Nucleosynthesis.\,\,We impose $\Delta N_\tf{eff} \!< 3.5$ as an exclusion bound.\,\,On the other hand, there exists an inconsistency between the Hubble constant observed today from the Cosmic Microwave Background (CMB) and the value measured from the celestial sources~\cite{Escudero:2019gzq,DiValentino:2021izs}.
To relieve this tension, $\Delta N_\tf{eff} \!> 0.2$ is preferred~\cite{Escudero:2019gzq}.\,\,Taking into account the constraints from BOREXINO~\cite{Kamada:2015era,Kaneta:2016uyt,Gninenko:2020xys}, CCFR~\cite{CCFR:1991lpl,Altmannshofer:2014pba}, and NA64~\cite{NA64:2024klw}, we identify a benchmark point (BP) $(m_{Z'}^{}, g^{}_X) = (11.5\,\tx{MeV}, 5 \times 10^{-4})$ as our numerical inputs, from which $\Delta N_\tf{eff} \simeq 0.3$, thereby relaxing the Hubble tension~\cite{Escudero:2019gzq}.\,\,Also, $\upsilon^{}_\Phi \sim m_{Z'}^{}/g^{}_X \sim {\cal O}(10)\,\tx{GeV}$.

\section{Higgs invisible decay}
With the dark sector, the SM-like Higgs boson $H_2$ has additional decay processes\,:\,$H_2 \to H_1 H_1, Z' Z',$ and $X X^\dagger$ if kinematically open.
These decay channels lead to invisible or non-standard decays of the SM-like Higgs boson, which is strongly bounded by the LHC data, ${\cal B}({}^{}H_2 \to \tx{Inv.}) < 0.13$~\cite{ParticleDataGroup:2022pth}. 
It is easy to check that the $H_2$ mainly decays into $H_1\!$ and $Z'\!$ pairs if $\lambda_{\Phi\!X} \!\lesssim\! {\cal O}(1)$, where $\Gamma_{H_2 \to H_1\!H_1} \!\simeq \Gamma_{H_2 \to Z'\!Z'} \!\propto \sin^2\!\theta \, m_{H_2}^3/\upsilon_\Phi^2 \gg \Gamma_{H_2 \to X\!X^\dagger} \!\propto \sin^2\!\theta {}^{} \lambda_{\Phi\!X}^2 \upsilon_\Phi^2/m^{}_{H_2}$. 
Hence, the constraint from the Higgs invisible decay on the $\sin\theta$ can be insensitive to $\lambda^{}_{\Phi\!X}\!$ as long as $\lambda^{}_{\Phi\!X}\!$ is small enough. 
Typically, the $\sin\theta$ should be less than $\sim\!10^{-2}$ to satisfy the Higgs invisible decay constraint. 
Note that the produced dark Higgs from the SM Higgs decay can further decay into the SM fermions, $f$. 
However, these decay channels are highly suppressed thanks to small $\sin\theta$ and yukawa couplings, where $\Gamma_{H_1 \to X\!X^\dagger} \!\propto \lambda^2_{\Phi\!X} \upsilon_\Phi^2 /m^{}_{H_1}  \gg \Gamma_{H_1 \to Z'\!Z'}\!\propto m^3_{H_1}/\upsilon_\Phi^2 \gg \Gamma_{H_1 \to f\bar{f}} \propto \sin^2\! \theta \, m^2_f m^{}_{H_1} /\upsilon_H^2$ in our interesting parameter space. 
Thus, all of the additional decay channels of the SM Higgs are invisible. 
It follows that the decay process $B^0 \!\to K^{\ast 0} H_1 \to K^{\ast 0} \mu^+\!\mu^-$ is also suppressed~\cite{Ovchynnikov:2023von}.
If the Belle II result on $B^+ \to K^+ \nu \bar{\nu}$ is confirmed near the future, light dark Higgs scenario can be confirmed by Higgs invisible decay channels.
The future expected bounds on Higgs invisible by HL-LHC \cite{Cepeda:2019klc}, ILC \cite{Potter:2022shg}, and FCCee \cite{Blondel:2021ema} will be $2.25\%,~0.16\%$, and $ 0.19\% $, respectively.
In top pannel of Fig.~\ref{Belle}, we present the future bounds on Higgs invisible decay.

\begin{figure}[t!]
\begin{center}
\hs{0.1cm}
\includegraphics[width=0.12\textwidth]{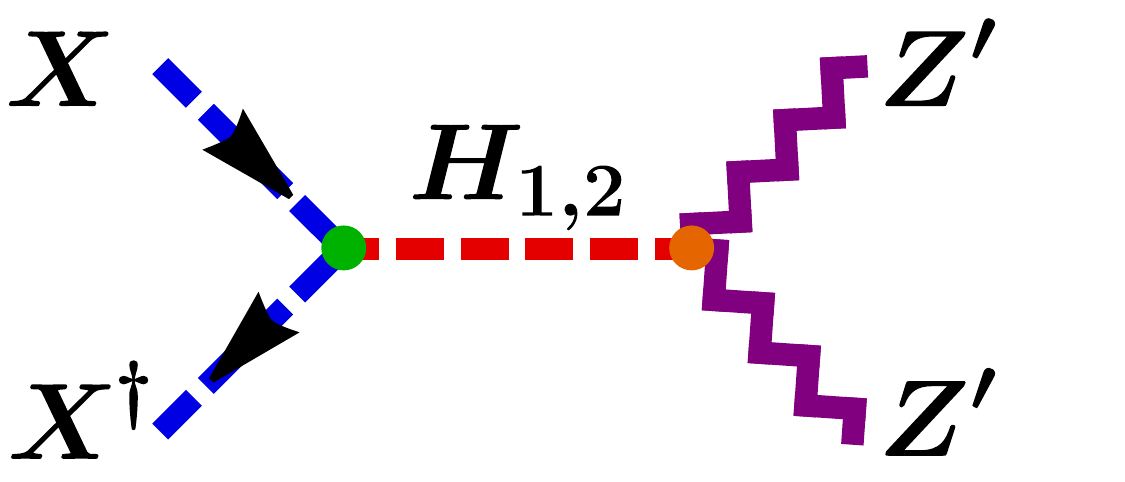}
\hs{-0.3cm}
\includegraphics[width=0.12\textwidth]{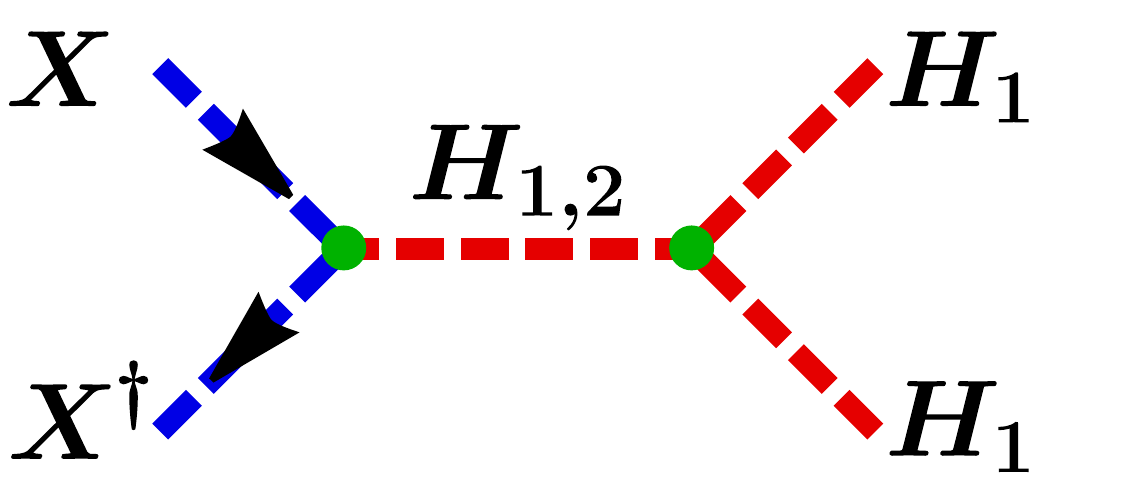}
\hs{-0.3cm}
\includegraphics[width=0.12\textwidth]{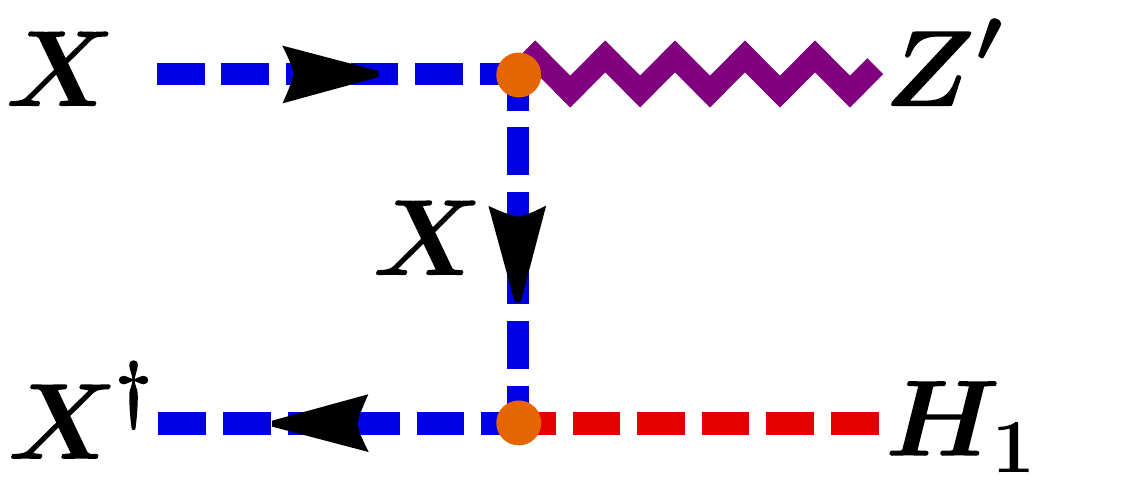}
\hs{-0.3cm}
\includegraphics[width=0.12\textwidth]{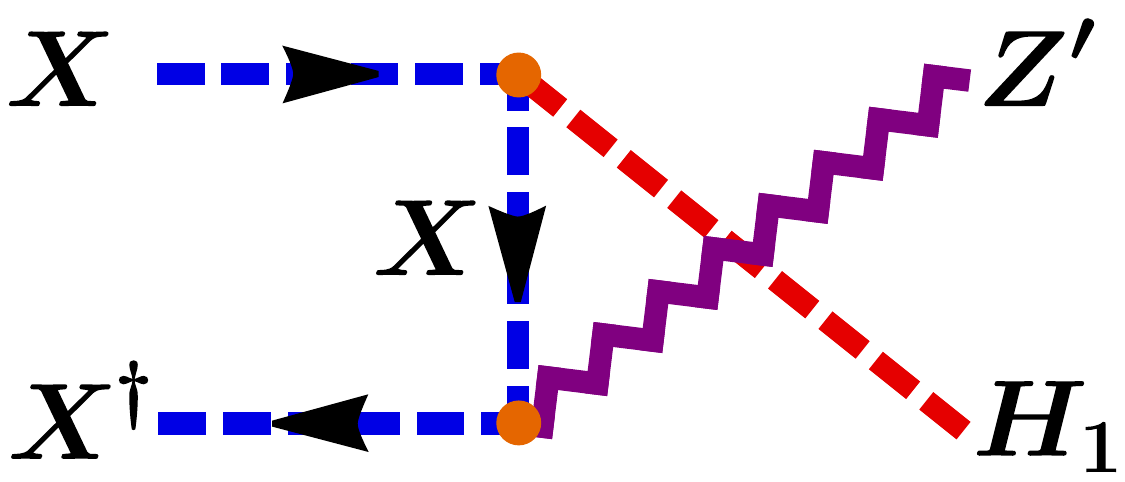}
\vs{-0.2cm}
\caption{Dominant Feynman graphs for complex scalar DM annihilating into the $Z'\!$ bosons and $H^{}_1\!$ bosons, where the arrow represents the $\tf{L}_\mu \! - \tf{L}_\tau \!$ charge flow.} \label{diagrams}
\end{center}
\vs{-0.5cm}
\end{figure}

\section{Dark Matter}
Let us now discuss the production of thermal freeze-out DM.
To get the right amount of DM relic density, it is required that the total DM annihilation cross section $\langle \sigma v \rangle$ for symmetric DM, $\Omega^{}_X = \Omega^{}_{X^\dagger}$, satisfying $\Omega^{}_\tf{DM} \hat{h}^2 \,= 0.1 \times (20\,\tx{TeV})^{-2}/\langle\sigma v\rangle$~\cite{Griest:1990kh}.

In $\UoneX$-charged DM models without a dark Higgs boson, the $\langle \sigma v \rangle$ only depends on $m^{}_{Z'}, g^{}_X$, and $m^{}_X$.\,\,To explain the $\Delta a_\mu$, $m^{}_{Z'}$ and $g^{}_X$ are fixed.\,\,Also, since the dominant DM annihilation process in these models is $X X^\dagger \!\to \nu\bar{\nu}$ via $s$-channel $Z'\!$ exchange, the correct DM relic abundance can only be produced when $m^{}_{Z'} \!\sim 2{}^{}m^{}_X$ \cite{Holst:2021lzm}.\,\,However, the authors in Ref.\,\cite{Baek:2022ozm} pointed out that this tight correlation between $m^{}_{Z'}\!$ and $m^{}_X\!$ can be completely bypassed if we include a dark Higgs boson, which plays an important role not only in the generation of dark photon mass but also in opening new DM annihilation channels and related processes.

We depict in Fig.\,\ref{diagrams} the dominant Feynman diagrams of the DM annihilation processes determining the relic abundance of DM in this model.\,\,They are $XX^\dagger \!\to Z'Z', H_1 Z', H_1H_1$, where the latter two processes are important only when $m^{}_X \!\gtrsim m^{}_{H_1}\!$.\,\,The processes $XX^\dagger \!\to \ell^+\ell^-, \nu^{}_\ell{}^{}\bar{\nu}^{}_\ell$ are suppressed by $\sin \theta$ and $g^{}_X$ for $\ell = \mu, \tau$, respectively, and by an additional kinetic mixing for $\ell = e$.\,\,Also, the $XX^\dagger \!\to Z'Z'$ process is governed by the $s$-channel $H_1\!$ exchange diagram (see the left diagram in Fig.\,\ref{diagrams}) because of the smallness of $g_X^4\!$. 
Thus, the total DM annihilation cross section in the $\sin\theta \ll 1$ and $\cos \theta \simeq 1$ limit (adopting hereafter) is approximately given by
\begin{eqnarray}
	\langle \sigma v \rangle 
	\simeq
	\frac{\lambda_{\Phi\!X}^2}{16{}^{}\pi{}^{}m_X^2}
	\frac{4{}^{}m_X^4 - 4{}^{}m_X^2 m_{Z'}^2 + 3{}^{}m_{Z'}^4}
	{\left(m_{H_1}^2 \!- 4{}^{}m_X^2 \right)\raisebox{0.5pt}{$\hs{-0.08cm}^{2}$} + 
		\Gamma_{H_1}^2 m_{H_1}^2} 
	\sqrt{1-\frac{m_{Z'}^2}{m_{X}^2}}
	~,
	\label{XX:ZpZp}
	\nonumber\\
\end{eqnarray}
where the decay width of $H_1\!$ is described as
\begin{eqnarray}
	\Gamma^{}_{H_1} 
	\simeq 
	\frac{\lambda_{\Phi\!X}^2 {}^{} \upsilon_\Phi^2}
	{16{}^{} \pi{}^{}m^{}_{H_1}}
	\sqrt{1 - \frac{4{}^{}m_X^2}{m_{H_1}^2}}
	~.
\end{eqnarray}
Notice that the dark Higgs can also decay into a pair of either $Z'\!$ or the SM charged particle.\,\,However, again due to the small values of $\sin\theta$ and $g^{}_X$, these decay channels are negligible.

 In $\UoneX\!$ DM models without a dark Higgs boson, DM has to be light, $m^{}_X \sim m^{}_{Z'}/2 \sim {\cal O}(10)\,\tx{MeV}$.\,\,The elastic scattering cross section of such light DM with the proton through the $Z'$ mediator is $\sigma^{X\tx{-}p}_\tf{el} \sim 10^{-46}{\rm cm^2}$, which is insensitive to the ongoing direct detection experiments~\cite{Holst:2021lzm}.

In Ref.~\cite{Holst:2021lzm}, they assumed that DM mass is around half of $Z'$ boson mass.
Therefore, the DM-nucleon elastic scattering cross section is much below the current scattering bound.
However, in our model, we can have a sizable DM-nucleon scattering thanks to the dark Higgs boson.
Our model does not require a close mass correlation between the $Z'$ boson and DM.
There are two contributions in the elastic scattering process.
The first one comes from Higgs contributions. 
If the $Z'$ contibution is negligible, the DM-nucleon scattering induced by the SM and dark Higgs boson exchanges is given by
\begin{align}
\sigma^{X\tx{-}n}_\tf{el} &\simeq 
\frac{4{}^{}\mu_n^2 {}^{} f_n^2 {}^{} \lambda^2_{\Phi X}  \sin^2  \theta}{\pi} 
\left( \frac{m^{}_n}{m^{}_X} \right)^{\hs{-0.12cm}2} 
\left( \frac{\upsilon^{}_\Phi}{\upsilon^{}_H} \right)^{\hs{-0.12cm}2} 
\left( \frac{1}{m_{H_1}^2} - \frac{1}{m_{H_2}^2} \right)^{\hs{-0.12cm}2}
\end{align}
where $f^{}_n=0.326$~\cite{Young:2009zb,Crivellin:2013ipa}, $m^{}_n$ is nucleon mass, and $\mu^{}_n = m^{}_X m^{}_n/(m^{}_X + m^{}_n)$ is the reduced mass of nucleon.
However, in some parameter space in our BP scenario, the scalar DM-nucleon scattering mediated by the light $Z'$ boson can have sizable due to the light $Z'$ and relatively heavy DM. 
The DM-nucleon scattering is given by
\begin{align}
\sigma^{X\tx{-}n}_\tf{el}  &\simeq \frac{\mu^2_n}{\pi}\frac{e^2 g^2_X Z^2 \epsilon^2 }{ A^2 m^4_{Z'} },
\end{align}
where $A$ and $Z$ are the number of proton and the nucleus, respectively.
Considering the current DM direct detection limit, DM mass less than $1.2$ GeV is allowed. 

If DM annihilates into charged SM particles during the CMB epoch, $T\sim {\rm eV}$, these SM particles can ionize and heat the surrounding gas. This spoils the evolution of the ionization history, altering the temperature and polarization anisotropies of the CMB ~\cite{Slatyer:2015jla}.
To avoid the CMB constraints on the low DM mass region, asymmetric DM~\cite{Kaplan:2009ag}, $p$-wave annihilation~\cite{Griest:1990kh}, or forbidden channel~\cite{DAgnolo:2015ujb} have been proposed.
In our DM model, the dominant DM annihilation channels are $XX^\dagger \!\to Z'Z', H_1H_1$.\,\,These channels are $s$-wave annihilation.\,\,However, the dark Higgs boson eventually decays into a DM pair, $H_1 \!\to X X^\dagger$, and the dark photon would decay into the active neutrinos, $Z' \!\to \nu\bar{\nu}$ since $m^{}_{Z'} \!< m^{}_\mu$.
During CMB epoch, the population of neutrinos might increase through DM annihilation because scalar DM annihilates to a pair of $Z'$ which subsequently decays to a pair of neutrinos.  
Considering the modification of $N_\tf{eff}$ via light DM $s$-wave annihilation to neutrinos, complex scalar DM mass below $8.2\,\tx{MeV}$ is disfavored~\cite{Chu:2023jyb}.\,\,In our study, we concentrate on $m^{}_X \gtrsim 20\,\tx{MeV}$.
Hence, we can naturally escape the CMB bound.
The impact of increase in $\Delta N_\tf{eff}$ is negligible and cannot be tested by CMB-stage 4  \cite{Escudero:2018mvt, Chu:2023jyb}. 

\section{Two- or three-body decays at Belle II}
The deviation reported by Belle II implies the presence of a BSM effect on the $b \to\! s \nu\bar{\nu}$ transition, which is treated as the missing energy $\slashed{E}$ in the final states of the $B^+\!$ to $K^+\!$ decay with a branching fraction of
\begin{eqnarray}
{\cal B}\!\left( B^+ \!\to K^+ \!\slashed{E} \,\right)_\tf{NP} 
=\, \left(1.8 \pm 0.7 \right) \times 10^{-5}
~.
\end{eqnarray}
This discrepancy has been interpreted by effective theories~\cite{Bause:2023mfe,Athron:2023hmz,
Allwicher:2023xba,He:2023bnk}, three-body $B \to K \chi\bar{\chi}$ decay with light DM or dark photon~\cite{Berezhnoy:2023rxx,Datta:2023iln,Altmannshofer:2023hkn,McKeen:2023uzo,
Fridell:2023ssf,Cheung:2024oxh}, and the other approaches~\cite{Felkl:2023ayn, Wang:2023trd}.
	
In Ref.\,\cite{Altmannshofer:2023hkn}, the authors stressed that the Belle II analysis provides information on the $q^2_{\rm rec}$ spectrum, indicating that there is a peak localized around $q^2_{\rm rec} = 4\,{\rm GeV^2}$.\,\,This can be described by a dark boson with a mass of 2\,GeV provided that its 
couplings to the SM fermions are sufficiently tiny.
In Ref.\,\cite{Altmannshofer:2023hkn}, the authors also pointed out that including BaBar data, the fit to the Belle II results in ${\cal B}(B^+ \to K^+ + \chi) = (5.1 \pm 2.1) \times 10^{-6}$, where $\chi$ is a new hidden particle.
The significance reduces $2.4\sigma$ for two body decay case.
Therefore, we will take two different branching ratios between 2-body decay and 3-body decay cases.
In comparison to Ref.\,\cite{Altmannshofer:2023hkn}, Ref.\,\cite{McKeen:2023uzo} studied a three-body decay involving a DM pair in a Higgs portal DM model.\,\,However, they predicted superabundant DM relic density.\,\,To obtain the correct relic abundance, they have to either introduce a new DM annihilation channel or modify the standard cosmology.\,\,Nonetheless, Ref.\,\cite{Fridell:2023ssf} performed likelihood analyses and concluded that the $B$ three-body decay is most favored.\,\,Combining all of these interpretations, the observed enhancement of the signals might indicate that new particles in the dark sector leave their imprints.
	
In Ref.\,\cite{Bird:2004ts}, they first proposed a bound on DM mass and its coupling using the $B^+ \to K^+ \nu\bar{\nu}$ decay channel in BaBar.\,\,They considered a real singlet scalar DM $(S)$ with the SM Higgs portal.\,\,However, the low mass region of this simple DM model is already ruled out by the CMB constraints since the relic density of DM is determined by the $s$-wave annihilation $SS \to H^{(\ast)}_2 \to f\bar{f}$.

In comparison with Ref.\,\cite{Bird:2004ts}, owing to the dark Higgs boson, the $B^+\!$ meson can go through a two-body decay, $B^+ \!\to K^+ \!H_1$ when $m^{}_{B^+} - m^{}_{K^+} > m^{}_{H_1}\!$ in our model.\,\,The corresponding two-body decay rate is
\begin{eqnarray}
	\Gamma^{}_{B^+ \to K^+\!H_1} 
	&\simeq&
	\frac{|\kappa^{}_{cb}|^2 \sin^2\!\theta}
	{64{}^{}\pi{}^{}m_{B^+}^3} \big[f^{}_0(m^2_{H_1})\big]^2 \!
	\left(\frac{m_{B^{+}}^2 - m_{K^+}^2}{m^{}_b - m^{}_s} \!\right)^{\hs{-0.1cm}2}
	\nonumber\\
	&&\times\sqrt{{\cal K}\big(m_{B^+}^2, m_{K^+}^2, m_{H_1}^2\big)}
	~,
\end{eqnarray}
where $\kappa^{}_{cb} \simeq 6.7 \times10^{-6}$ is the one-loop vertex correction after integrating out the top quark and $W$ boson, $f^{}_0(q^2)$ with $q^2 = (\,p^{}_B - p^{}_K)^2$ is the $B^- \to K^-$ transition form factor which can be found in Ref.\,\cite{Parrott:2022rgu}, and ${\cal K}(a,b,c) \equiv a^2 + b^2 + c^2 - 2 \big(ab + bc + ac\big)$.

\begin{figure}
\centering
\includegraphics[width=0.98\linewidth]{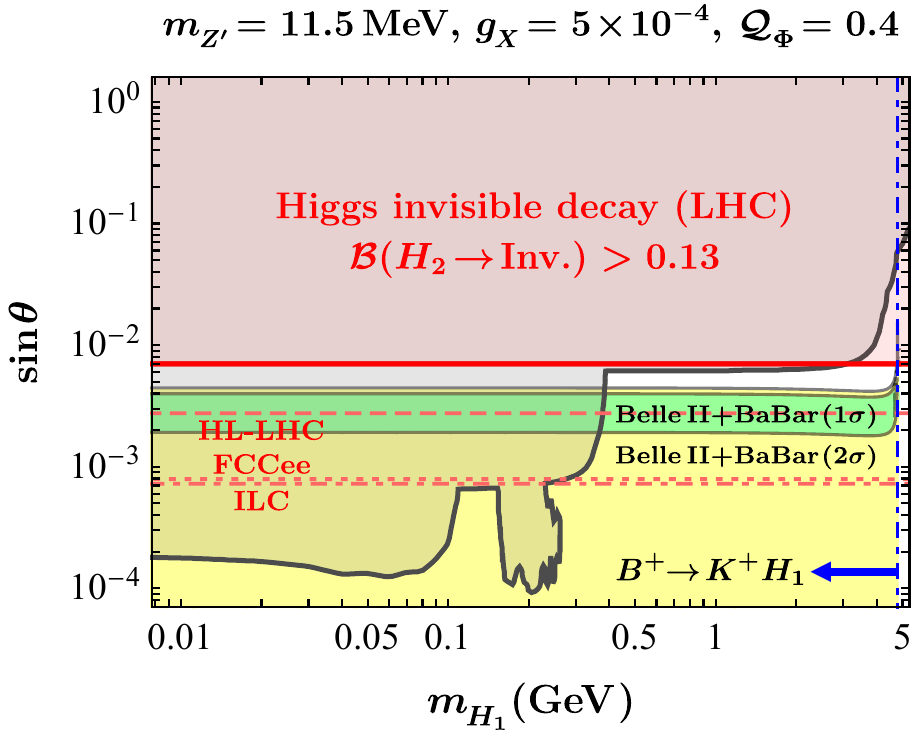}
\includegraphics[width=0.98\linewidth]{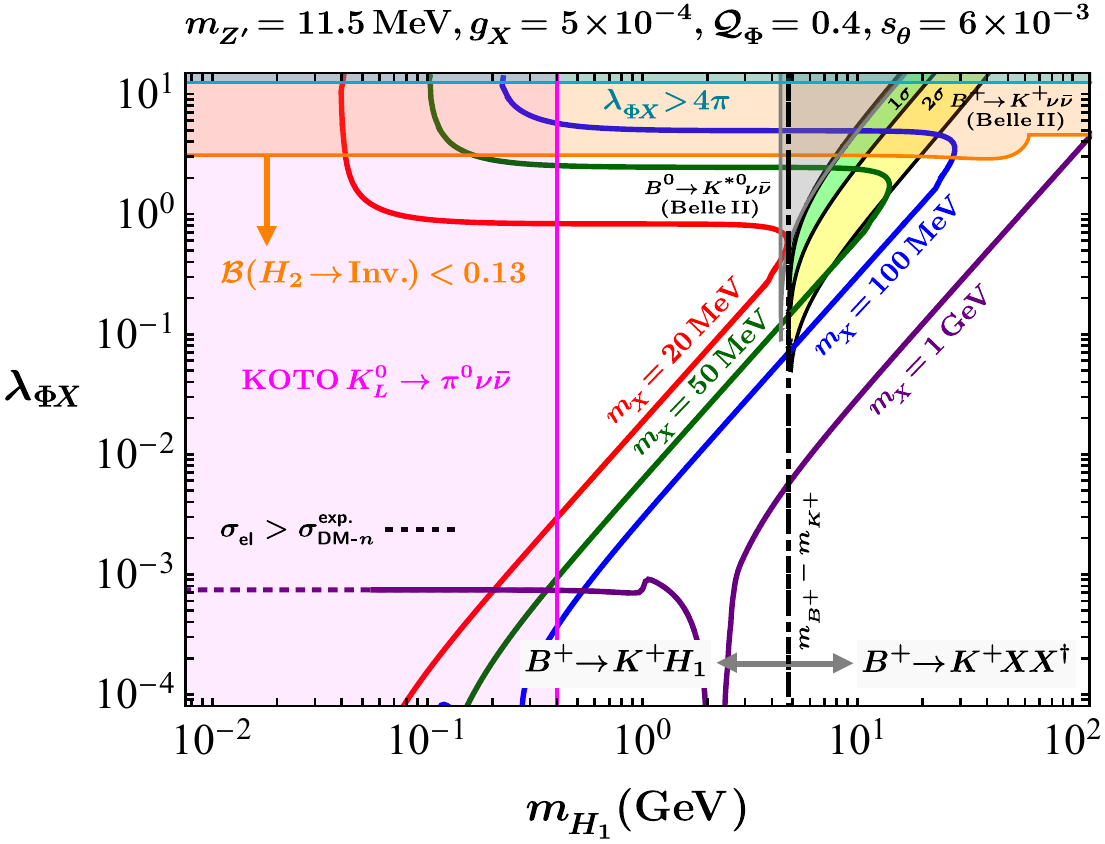}
\vs{-0.3cm}
\caption{(Top) Allowed parameter space of $(\sin \theta, m^{}_{H_1})$ plane. The region inside green (yellow) is allowed at $1{}^{}\sigma\,(2{}^{}\sigma)$ C.L. by Belle II $B^+ \to K^+ \!H_1$, where $H_1$ decays to either DM or $Z'$. The gray area is excluded by various experiments from Belle $B^0 \to K^{\ast 0} \nu\bar{\nu}$, and $K^+\to \pi^+ {\rm inv.}$. (Bottom) The preferred parameter space in the $(m^{}_{H_1}, \lambda^{}_{\Phi\!X})$ plane.\,\,The region inside green (yellow) is allowed at $1{}^{}\sigma\,(2{}^{}\sigma)$ C.L. by Belle II $B^+ \!\!\to K^+\!XX^\dagger$.\,\,The orange area is ruled out by the SM Higgs invisible decay at LHC~\cite{ParticleDataGroup:2022pth}.\,\,The solid lines correspond to the thermal freeze-out DM relic density which is consistent with the Planck observation, $\Omega^{}_\tf{DM} \hat{h}^2 = 0.12$~\cite{Planck:2018vyg}.\,\,The dashed lines are disfavored by the DM direct detection bound from XENONnT~\cite{XENON:2023cxc}.\,\,The gray region is excluded by the $B^0 \to K^{\ast 0} \nu\bar{\nu}$ bound.} 
\label{Belle}
\vs{-0.3cm}
\end{figure}

Similar to Ref.\,\cite{Bird:2004ts}, when the dark Higgs becomes off-shell, where $m^{}_{H_1} > m^{}_{B^+} - m^{}_{K^+} > 2{}^{}m^{}_X$, the $B^+\!$ meson can have a three-body decay, $B^+ \!\to K^+\! H_1^{(\ast)} \!\to K^+\!X X^\dagger$. The corresponding three-body decay rate is
\begin{eqnarray}
	\Gamma^{}_{B^+ \to K^+\!X\!X^\dagger} 
	&\simeq & 
	\frac{\lambda_{\Phi\!X}^2 {}^{} \upsilon_\Phi^2 \, |\kappa^{}_{cb}|^2 \sin^2\!\theta
	}{1024{}^{}\pi^3{}^{}m_{B^+}^3} \!
	\left(\frac{m_{B^+}^2 - m_{K^+}^2}{m^{}_b - m^{}_s} \!\right)^{\hs{-0.1cm}2} \!
	\nonumber\\
	&& 
	\times\mathop{\mathlarger{\int}} \!\! \dd q^2
	\frac{\mathcal{I}(q^2) \big[f^{}_0(q^2)\big]^2 \big(m^2_{H_1} - m^2_{H_2}\big)^{\hs{-0.03cm}2}}
	{\big(q^2-m^2_{H_1}\big)\raisebox{0.5pt}{$^2$}
		\big(q^2-m^2_{H_2}\big)\raisebox{0.5pt}{$^2$}}
	~,
	\nonumber\\
\end{eqnarray}
where $4{}^{}m_X^2 \leq q^2 \leq (m^{}_{B^+} - m^{}_{K^+})^2$, and 
\begin{eqnarray}
	{\cal I}(q^2)
	\,=\,
	\sqrt{1 -\frac{4{}^{}m_X^2}{q^2}} 
	\sqrt{{\cal K}\big(m_{B^+}^2, m_{K^+}^2, q^2\big)}
	~. 
\end{eqnarray}
Note that the $B^+ \!\to K^+\!Z'Z'$ decay is also open, but its contribution is negligible as
$\Gamma_{B^+ \!\to K^+\!Z'\!Z'}/\Gamma_{B^+ \!\to K^+\!X\!X^\dagger} \!\propto m^4_B/ (\lambda^{2}_{\Phi\!X} \upsilon_\Phi^4) \lesssim 7\times 10^{-3}$, where $\lambda_{\Phi\!X}^{} \gtrsim {\cal O}(0.1)$ to explain the observed DM relic density and $\BKnunu$ excess (see the bottom panel of Fig.\,\ref{Belle}).

\section{Numerical results}

We present in the top panel of Fig.\,\ref{Belle} the mixing angle versus the dark Higgs mass with our BP, where the gray region is ruled out by Belle II $B^0 \to K^{\ast 0} \nu\bar{\nu}$ channel~\cite{Belle:2017oht}, KOTO~\cite{KOTO:2020prk}, and $K^+ \to \pi^+ + {\rm invisibles}$ from NA62~\cite{NA62:2021zjw} and the red-shaded region is excluded by the Higgs invisible decay constraint~\cite{ParticleDataGroup:2022pth}.\,\,The $\BKnunu$ excess at Belle II is explained by the two-body decay channel $B^+ \!\to K^+\!H_1$ in the green and yellow bands.\,\,According to Ref.\,\cite{Altmannshofer:2023hkn}, 2\,GeV new light particle is preferred because of $q^2\!$ spectrum information provided by Belle II.
For two-body case, the mixing angle $\sin\theta < 4.5 \times 10^{-3}$ is preferred due to BaBar and Belle II.
For three-body case, $\sin\theta \leq 7 \times 10^{-3}$ is allowed considering the Higgs invisible decay limit.
We will take $\sin\theta = 6 \times 10^{-3}$ as our input in the three-body decay at Belle II.

In the bottom panel of Fig.\,\ref{Belle}, we show the parameter region of $\lambda^{}_{\Phi\!X}\!$ varying with $m^{}_{H_1}$, where the color curves with different DM masses satisfy the relic abundance of DM.\,\,The orange- and cyan-shaded regions are excluded by the Higgs invisible decay constraint and perturbativity, respectively.\,\,On the left side of the dash-dotted line, where $m^{}_{H_1} \!= m^{}_{B^+} - m^{}_{K^+}$, the $\BKnunu$ excess can be explained by the $B^+ \!\to K^+\!H_1$ when $m^{}_{H_1} \!\gtrsim 0.4\,\tx{GeV}$ based on the top panel of Fig.\,\ref{Belle}.\,\,Notice that $m^{}_X \gtrsim 1$ GeV is not allowed due to the DM direct detections.\,\,On the right side of the dash-dotted line, the $\BKnunu$ excess is explained within the green and yellow regions by the three-body decay channel $B^+ \!\to K^+\!XX^\dagger$.\,\,The bounds from ${\cal B}(B^0 \to K^{\ast 0} \nu\bar{\nu}) < 1.8 \times 10^{-5}$ (gray region)~\cite{Belle:2017oht} partially excludes the preferred region for Belle II excess.\,\,Note that these regions are only valid for the light DM masses, where $m^{}_X \ll m^{}_{B^+}$.

The flat and diagonal behavior of the color curves in the low- and high- $H_1\!$ mass regions is simple to understand.\,\,This is because $\langle \sigma v \rangle \propto \lambda_{\Phi\!X}^2/m_X^2\,\big(\lambda_{\Phi\!X}^2 m_X^2/m_{H_1}^4\big)$ for $m^{}_{H_1} \!\ll m^{}_X\,(m^{}_{H_1} \!\gg m^{}_X)$ and $\langle \sigma v \rangle$ is nearly a constant.\,\,However, these color curves become flat again when $\lambda^{}_{\Phi\!X}$ gets much larger.\,\,This unexpected behavior originates from the $H_1\!$ decay width in the denominator of the $X X^\dagger \!\to Z'Z'$ cross section, where $\Gamma_{H_1}^2 m_{H_1}^2 \!\propto \lambda_{\Phi\!X}^4\upsilon_\Phi^4$ is dominant over $m_{H_1}^4\!$ such that $\langle \sigma v \rangle \propto m_X^2/(\lambda_{\Phi\!X}^2 \upsilon_\Phi^4)$.
The inflection point is around $m^{}_{H_1} \!\!\sim m^{}_X/(\langle \sigma v \rangle^{1/2} \upsilon^{}_\Phi)$.\,\,After this inflection point, the $H_1\!$ decay width enters the phase space suppression when $m^{}_{H_1} \!\gtrsim 2{}^{}m^{}_X$, and then the required $\lambda^{}_{\Phi\!X}\!$ to fit the DM relic density becomes even larger.\,\,Due to this interesting behavior, the light DM mass between $20~\tx{MeV} \lesssim m^{}_X \lesssim 60\,\tx{MeV} $ is allowed.

\section{conclusions}
After the Belle II excess announcement, a number of literature tried without success to explain both the observed DM relic density and the enhancement of Belle II simultaneously. That is, fitting the Belle II excess with light DM provides an extremely large relic abundance.
To reproduce the correct relic density, either introducing new DM annihilation channels or allowing DM to decay was necessary.

In this letter, we have considered the simplest UV-complete gauged $\UoneX$-charged complex scalar DM model.\,\,We have found that the dark Higgs boson mass $0.4\,\tx{GeV} \lesssim m^{}_{H_1} \lesssim 10\,\tx{GeV}$ with the upper (lower) bound coming from the $\BKnunu$ excess 
(KOTO if $\sin\theta \sim 6 \times 10^{-3}$), the complex scalar DM mass $10\,\tx{MeV} \lesssim m^{}_X \lesssim 1.2~ \tx{GeV}$ with the upper (lower) bound coming from the direct detection experiments ($\Delta N^{}_\tf{eff}$), and the dark photon mass $m^{}_{Z'} \sim 10\,\tx{MeV}$ with $g^{}_X \sim 5 \times 10^{-4}$ to explain the muon $g-2$.\,\,Attributing to these light dark particles in this model, we can achieve the integrated solution of $\BKnunu$ at Belle II, $\Delta a^{}_\mu$, thermal DM relic density, and Hubble tension.\,\,Our analysis makes another case where the dark Higgs boson plays a crucial role in DM phenomenology.

In this scenario, $Z'$ is light enough to contribute to the charged pion decay at the level of 
$\Gamma ( \pi^+ \rightarrow \mu^+ \nu_\mu Z' )/ \Gamma (  \pi^+ \rightarrow \mu^+ \nu_\mu  ) 
\sim 3 \times 10^{-9}$, which is well below the current sensitivity of PIENU, ${\cal O}(10^{-5})$ \cite{PIENU:2021clt}.\,\,Our light dark particles can be fully tested by upcoming CMB stage-4 and updated Belle II and NA64 data in the future.\,\,Our scenario can also have nontrivial implications for large scale structure of the Universe through the measurements of matter power spectrum.\,\,Scattering between DM and dark radiation (neutrinos in our model) through the $t$-channel $Z'$ exchange can affect $\sigma^{}_8$ tension.

\bigskip
\begin{acknowledgments}
We are grateful to Josef Pradler, Michael Schmidt, Satoshi Shirai, and Fanrong Xu for useful discussions.\,\,This work is supported by KIAS Individual Grants under Grants No.\,PG081202 (SYH), PG021403 (PK), and by National Research Foundation of Korea (NRF) Research Grant NRF-2019R1A2C3005009 (JK, PK) and No. NRF-2022R1A2C2003567, No. RS-2024-00341419 (JK).
\end{acknowledgments}

\bibliographystyle{utphys}
\bibliography{literature}

\onecolumngrid
\appendix

\clearpage
\section{Appendix}
In the main body of this Letter, we focused on the BP where both the muon anomalous magnetic moment with $5.1\,\sigma$ deviation and Hubble tension can be resolved.
As a matter of fact, the muon anomalous magnetic moment has had discrepancy between the SM theoretical prediction and the experimental measurement for more than two decades if we consider the SM expectation from the value in the white paper \cite{Aoyama:2020ynm}.
On the other hand, the uncertainties in first-principles lattice calculations have now been refined to a level comparable to the data-driven calculation.
Recently, CMD-3 collaboration reported that new $e^+ e^- \to 2\pi$ production data in the energy range $E_{\rm CM} < 1$ GeV \cite{CMD-3:2023rfe}.
If the new CMD-3 result is used for the hadronic vacuum polarization,
$\Delta a_\mu$ is given by
\begin{align}
\Delta a_\mu &= \left(4.9 \pm 5.5 \right) \times 10^{-10}, 
\end{align}
which is consistent with the combined experimental data from BNL and Fermilab muon $g-2$.
The BMW collaboration recently announced that new results is only $0.9\sigma$ deviation from the experimental data.

If new CMD-3 and BMW data is correct, dark photon $Z'$ receives an additional 
bound from the muon $(g-2)$ experiments.
Here we take new benchmark point $(m_{Z'}=10{\rm MeV}, g_X=10^{-4})$ 
which gives $\Delta a_\mu = 10^{-10}$.
This new BP can substantially relax the Hubble tension and be consistent  with current $\Delta a_\mu$ data within $1\sigma$ C. L.
\begin{figure}[tbh]
\centering
\includegraphics[width=0.44\linewidth]{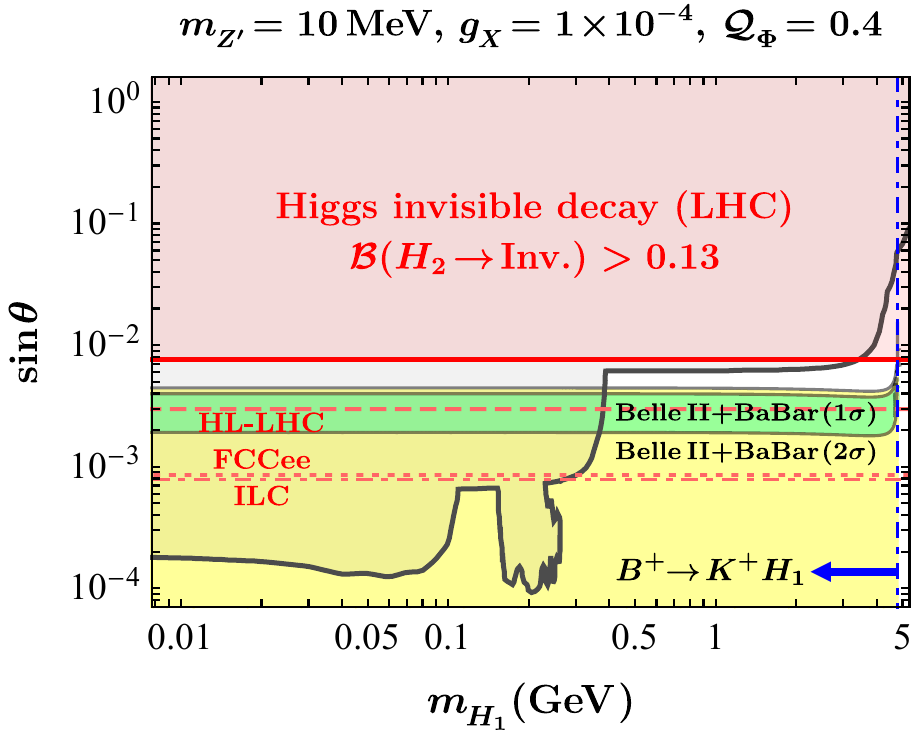}
\includegraphics[width=0.46\linewidth]{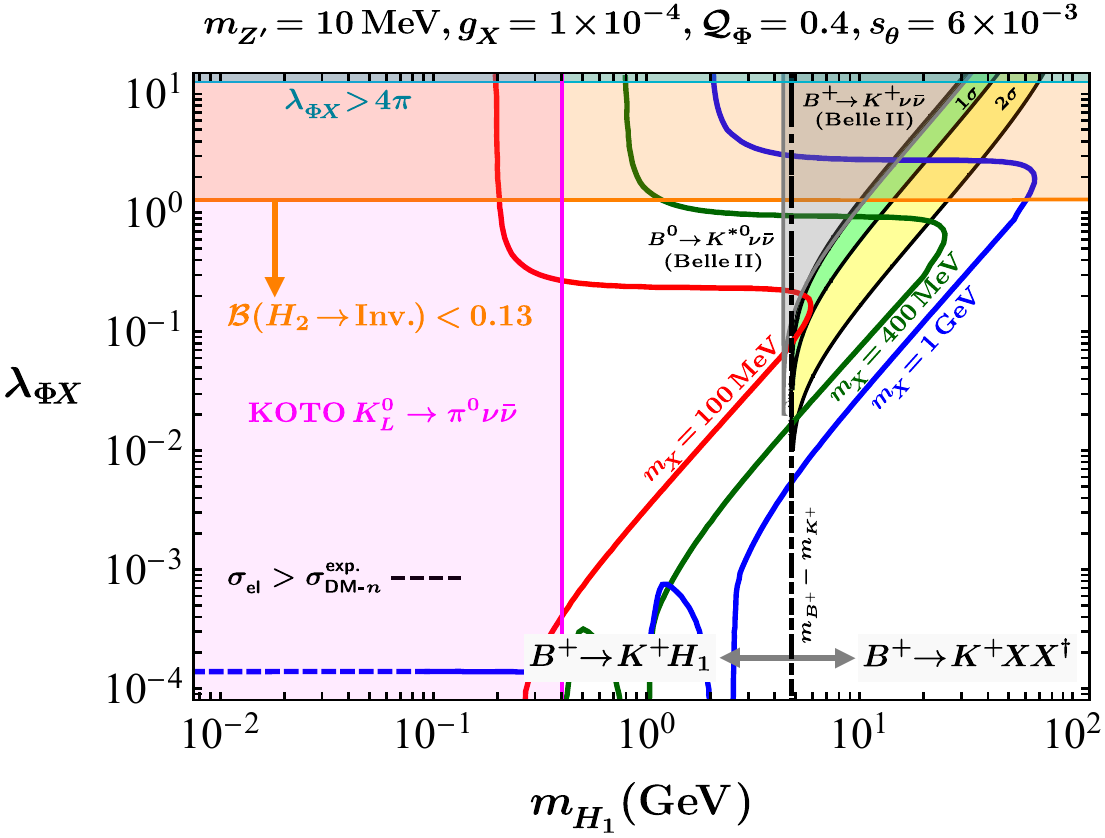}
\vs{-0.3cm}
\caption{(Left) Allowed parameter space of $(\sin \theta, m^{}_{H_1})$ plane. The region inside green (yellow) is allowed at $1{}^{}\sigma\,(2{}^{}\sigma)$ C.L. by two-body decay case in the Belle II. The gray area is constrained by diverse experiments from Belle $B^0 \to K^{\ast 0} \nu\bar{\nu}$, and $K^+\to \pi^+ {\rm inv.}$. (Right) The preferred parameter space in the $(m^{}_{H_1}, \lambda^{}_{\Phi\!X})$ plane.\,\,The region inside green (yellow) is allowed at $1{}^{}\sigma\,(2{}^{}\sigma)$ C.L. by three-body decay case in the Belle II. \,\,The orange area is ruled out by the SM Higgs invisible decay at LHC~\cite{ParticleDataGroup:2022pth}.\,\,The solid lines correspond to the thermal freeze-out DM relic density which is consistent with the Planck observation, $\Omega^{}_\tf{DM} \hat{h}^2 = 0.12$~\cite{Planck:2018vyg}.\,\,The dashed lines are disfavored by the DM direct detection bound from XENONnT~\cite{XENON:2023cxc}.\,\,The gray region is excluded by the $B^0 \to K^{\ast 0} \nu\bar{\nu}$ bound.} 
\label{Belle:NewBP}
\vs{-0.3cm}
\end{figure}

In the left panel of Fig.~\ref{Belle:NewBP}, we show the mixing angle versus the dark Higgs mass with our new BP.
The gray region is excluded by Belle II $B^0 \to K^{\ast 0} \nu\bar{\nu}$ channel~\cite{Belle:2017oht}, KOTO~\cite{KOTO:2020prk}, and $K^+ \to \pi^+ + {\rm invisibles}$ from NA62~\cite{NA62:2021zjw}.
The red-shaded region is excluded by the Higgs invisible decay constraint~\cite{ParticleDataGroup:2022pth}.
Following Ref.\,\cite{Altmannshofer:2023hkn}, 2\,GeV new light particle is favored because of $q^2_{\rm rec}$ spectrum data.
Compared to Fig.~\ref{Belle}, Higgs invisible decay bound is just slightly changed because new BP has larger VEV.
In the right panel of Fig.~\ref{Belle:NewBP}, we present the parameter region of $\lambda^{}_{\Phi\!X}\!$ varying with dark Higgs mass.
All of the solid lines satisfy the DM relic density.
The orange- and cyan-shaded regions are ruled out by the Higgs invisible decay constraint and perturbativity, respectively.
On the left side of the dash-dotted vertical line, where $m^{}_{H_1} \!= m^{}_{B^+} - m^{}_{K^+}$, the $\BKnunu$ excess can be explained by the $B^+ \!\to K^+\!H_1$.
On the right side of the dash-dotted line, the $\BKnunu$ excess can be resolved within the green ($1\sigma$) and yellow ($2\sigma$) regions by the three-body decay channel $B^+ \!\to K^+\!XX^\dagger$.
Thanks to larger VEV, the allowed DM mass lies between 90 MeV and $\sim 450$ MeV.

\end{document}